\documentclass[aps,showpacs,prl,twocolumn,amsmath,amssymb,letterpaper,floatfix]{revtex4}
\pdfoutput=1
\usepackage{graphicx}
\usepackage{tabularx}
\usepackage{wrapfig}
\usepackage{setspace}

\makeatletter
\renewcommand{\@makefntext}[1]{\parindent=1em\noindent\hbox to 1.8em{\hss$^{\@thefnmark}$}#1}
\renewcommand{\@footnotemark}{\hbox{\mathsurround=0pt$^{\@thefnmark}$}}

\makeatother

\begin{document}

\title{Symmetries of mesons after unbreaking of chiral symmetry
and their string interpretation}

\author{M. Denissenya}
\email{mikhail.denissenya@uni-graz.at}
\author{ L.Ya. Glozman }
\email{leonid.glozman@uni-graz.at}
\author{ C.B. Lang }
\email{christian.lang@uni-graz.at}
\affiliation{Institute of Physics,  University of Graz, A--8010 Graz, Austria}

\begin{abstract}
 Using the  chirally 
invariant overlap Dirac operator we remove its lowest-lying quasizero
modes
from the valence quark propagators and study evolution of 
isovector mesons with $J=1$. At the truncation level
about 50 MeV 
$SU(2)_L \times SU(2)_R$ and $U(1)_A$ symmetries get restored. However,
we observe a degeneracy
not only within the  chiral and $U(1)_A$ multiplets, but also a degeneracy
of all possible chiral multiplets, i.e.,  the observed
quantum levels have a symmetry larger than  $U(2)_L \times U(2)_R$  and 
their energy does not depend on the spin orientation of  quarks and their parities.
We offer a possible interpretation of these energy levels as the quantum
levels of the dynamical QCD string. The structure of the radial $J=1$
spectrum is compatible with $E =(n_r +1)\hbar\omega$ with
$\hbar\omega = 900 \pm 70$ MeV. 
\end{abstract}

\maketitle

\paragraph{1. Introduction.} A consistent and systematic picture of hadrons 
made
of light quarks is  missing. There is a general understanding that both
confinement and spontaneous breaking of chiral symmetry (SB$\chi$S) are 
important for hadronic mass generation. A large degeneracy is seen in the
highly excited mesons \cite{G3,G2}, which is however absent
in the observed spectrum of mesons with masses
below 1.7 - 1.8 GeV. The physics of the low-lying hadrons should
be affected by the SB$\chi$S that might obscure the primary confinement picture.
We want to disentangle the primary physics that is responsible for a 
genesis of hadrons from the SB$\chi$S.

In order to address this issue we  artificially remove the chiral
symmetry breaking dynamics from the valence quark propagators keeping at the
same time the gluonic gauge configurations intact \cite{LS,GLS,DG}. 
It has been known for a long time that the quark condensate of the vacuum is directly
related to the density of the close-to-zero modes of the Dirac operator \cite{BC}.
The quark propagator can be written in terms of projectors onto the
eigenmodes of the Dirac operator.
We subtract from the full propagator its lowest-lying modes, that represent
 only a tiny part of the full amount of modes,
  \begin{equation}\label{eq:RD}
  S_{RD(k)}=S_{Full}-
  \sum_{i=1}^{k}\,\frac{1}{\lambda_i}\,|\lambda_i\rangle \langle \lambda_i| \\, 
 \end{equation}
  \noindent
where $\lambda_i$ and $|\lambda_i\rangle$ are the  eigenvalues and
the corresponding eigenvectors of the 
Dirac operator. This way  we artificially restore ("unbreak") the chiral symmetry.

In ref. \cite{GLS} we have noticed that after unbreaking the chiral symmetry
a degeneracy of hadrons develops that is higher than $SU(2)_L \times SU(2)_R$.
Such a degeneracy should reflect the underlying symmetry and dynamics. Still 
the quality
of our data that support a degeneracy was not too high.

A reason for not a high quality was that in Refs. \cite{LS,GLS,DG}  
a not-chirally-symmetric
lattice Dirac operator of the Wilson type was used. An eigenvalue decomposition in 
terms of the Hermitian
Dirac operator $\gamma_5 D$ was adopted. It causes ambiguity as there is no 
direct bijective relation between the low-lying modes of  $\gamma_5 D$  and of the
Dirac operator $D$; exact zero modes  could not be exactly identified.
In the present work we use the manifestly chirally
invariant overlap Dirac operator \cite{OVERLAP} where such problems are absent.

From the truncated quark propagators we construct hadron propagators.  When we
observe an exponential decay signal of the corresponding correlation function we
interpret it as a physical state and extract its mass. Our task is to see what will
happen with hadrons and their masses when we subtract the lowest-lying modes
of the Dirac operator.

\paragraph{2. Lattice technology.} We adopt 100 gauge field configurations generated
with $n_f=2$ dynamical overlap fermions on a $16^3\times 32$ lattice with the spacing
$a \sim 0.12$ fm \cite{KEK,Aoki:2012pma}. The pion mass in this ensemble is 
$M_{\pi} =289(2)$ MeV \cite{Noaki:2008iy}.
For each hadron we use a set of up to eight interpolators with different Dirac structures
(see Table \ref{tab:ints}) 
and different exponential smearings of sources and sinks with several smearing widths. 
We determine correlators for all polarizations.
The quark propagators were generously provided by the JLQCD collaboration and were computed by combining of the exact   100 low modes with stochastic estimates for the higher modes \cite{Aoki:2012pma}.

For
the analysis of the cross-correlation matrices we employ the variational method
and solve the generalized eigenvalue problem \cite{VAR}. In the figures we show effective
energy values (nearest neighbour slopes)  to demonstrate plateau behaviour.
The horizontal bars on the effective energy plots 
 indicate the fit range.
The one-exponential fits are to the eigenvalues of the correlation matrices
in the range shown in the 
effective mass plots. The error bars are determined by the single-elimination jackknife method. Table \ref{tab:ams} gives
the resulting numbers for two stages of truncation, $k=10, 20$.

\begin{table}[t]
%    \begin{center}
 \caption{Interpolators for the $\rho(1^{--})$,  $a_1(1^{++})$ and $b_1(1^{+-})$ mesons 
  ($i,j =1,2,3$)  with 7 different
smearing widths at sink/source. }\label{tab:ints}
 \begin{ruledtabular}
   \begin{tabular}{ccc}
channel & interpolator types& no. of interpolators\\
    \hline
        $\rho$ & $\overline{q} \gamma_i \vec \tau q$,  $ \overline{q}
	\gamma_i\gamma_t \vec \tau q$&8\\
        $a_1$  & $\overline{q} \gamma_i \gamma_5 \vec \tau q$&5\\
        $b_1$  & $\overline{q} \gamma_i \gamma_j \vec \tau  q$&6 \\
    \end{tabular}
\end{ruledtabular}
%    \end{center}
  \end{table}
  
\paragraph{3. Observations.}
We have studied the ground state and the excited states of all possible $\bar q q$
isovector
$J=1$ mesons, i.e., $\rho (1^{--})$, $a_1 (1^{++})$, $b_1 (1^{+-})$.
Typical results for the eigenvalues of the correlation matrices and the effective
mass plots  are shown in Fig. 1. Upon
removal of the lowest eigenmodes of the Dirac operator from the valence
quark propagators we observe an improving signal for clean exponential decay.
The evolution of  masses of the ground and excited states are shown in Fig. \ref{fig:histo}.
Around a truncation energy of ~40-65 MeV (approximately 10 eigenmodes in our ensemble)
an onset of a  degeneracy of four 
states, $\rho, \rho', a_1,b_1$, as well as a degeneracy of their excited
states is seen
\footnote{Note, that in order to observe two degenerate
independent  orthogonal $\rho$-mesons it is necessary to have in the 
correlation matrix operators with two different chiral structures, see the
subsequent discussion. With the vector current type operator 
$\bar q  \gamma_{\mu} \vec \tau q$ alone one of these states will not be seen, 
as that second $\rho$-meson  does not couple to the vector current \cite{G2,G3}.}.
The degeneracy indicates a yet unknown symmetry. 
\begin{table}[t]
 \caption{Masses of the $\rho$, $\rho'$, $a_1$, $b_1$ mesons extracted at the specific truncation
levels $k$. }\label{tab:ams}
     \noindent 
\begin{ruledtabular}
\begin{tabular}{ccccc}
         & \multicolumn{2}{c}{$k=10$} & \multicolumn{2}{c}{$k=20$}   \\ 
    \hline
 $am$    & $n_r=0$ & $n_r=1$ & $n_r=0$ & $n_r=1$ \\ 
    \hline
        $\rho$ & 0.573(10) & 1.096(57)  & 0.625(7) & 1.107(58) \\
 	 $\rho'$ & 0.589(13) & 1.108(48)  & 0.614(8)  & 1.139(48) \\  
        $a_1$  & 0.576(10)   &	1.051(62) & 0.615(8)  & 1.062(61) \\  
         $b_1$  & 0.568(14)   &	1.077(51) & 0.607(8)  & 1.106(51) \\
    \end{tabular}
    \end{ruledtabular}
  \end{table}

\begin{figure*}[th]
  \centering
  \hspace*{12pt}
  \includegraphics[width=0.45\textwidth]{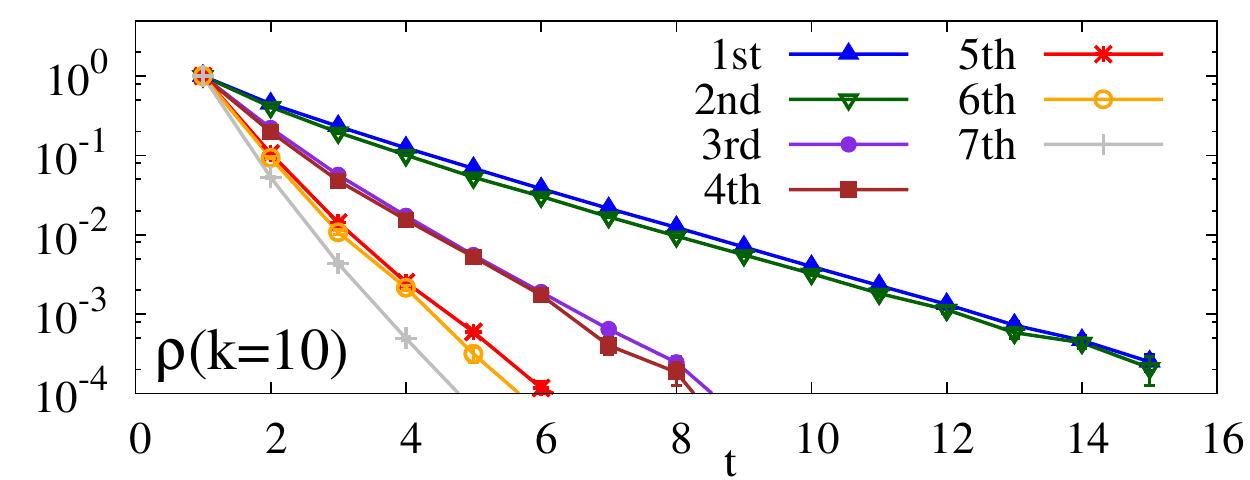}\hfill
  \includegraphics[width=0.45\textwidth]{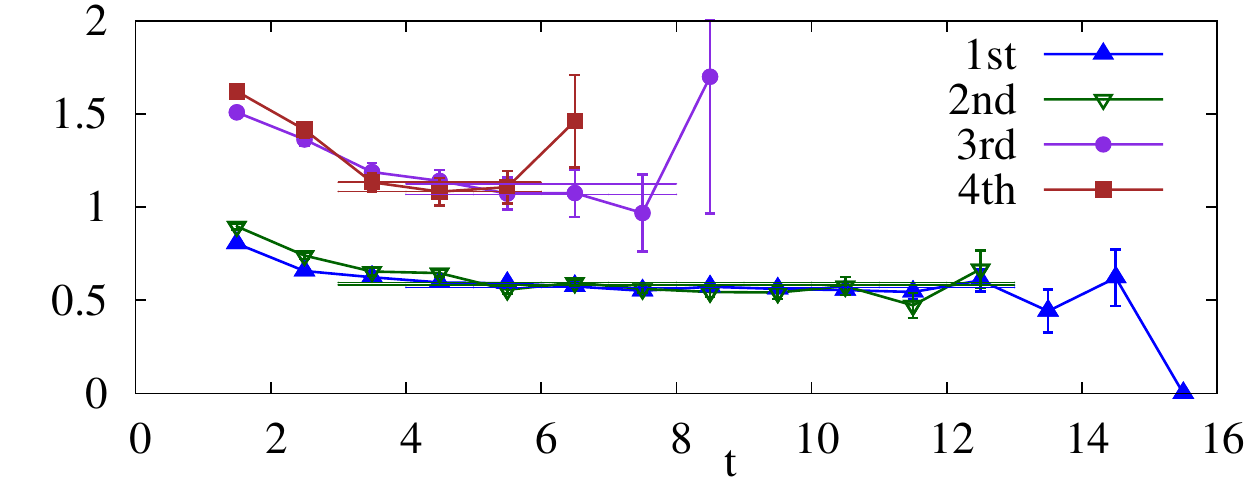}\hspace*{24pt}\hfil\\
  \hspace*{12pt}
  \includegraphics[width=0.45\textwidth]{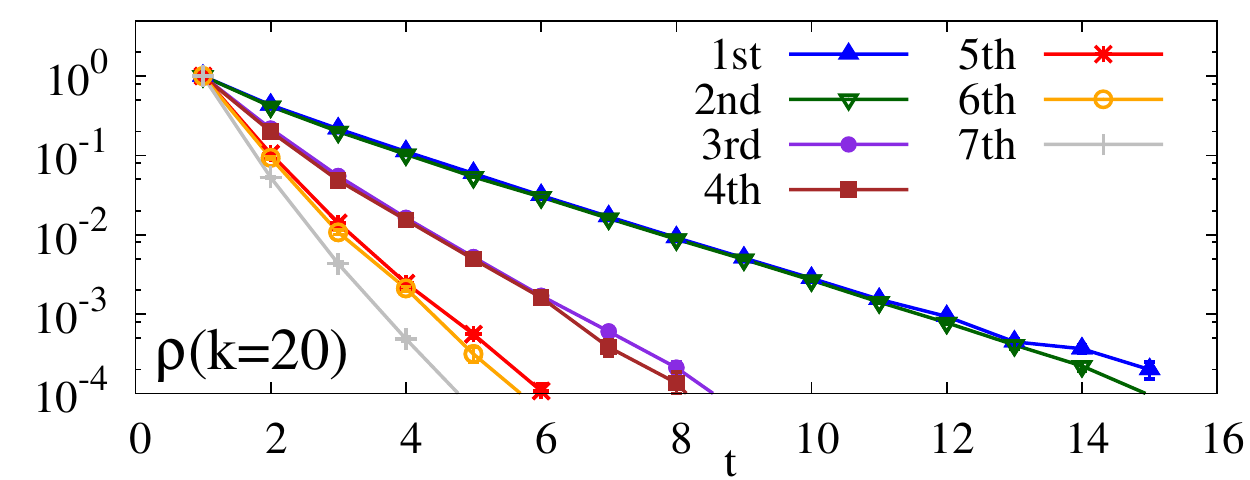}\hfill
  \includegraphics[width=0.45\textwidth]{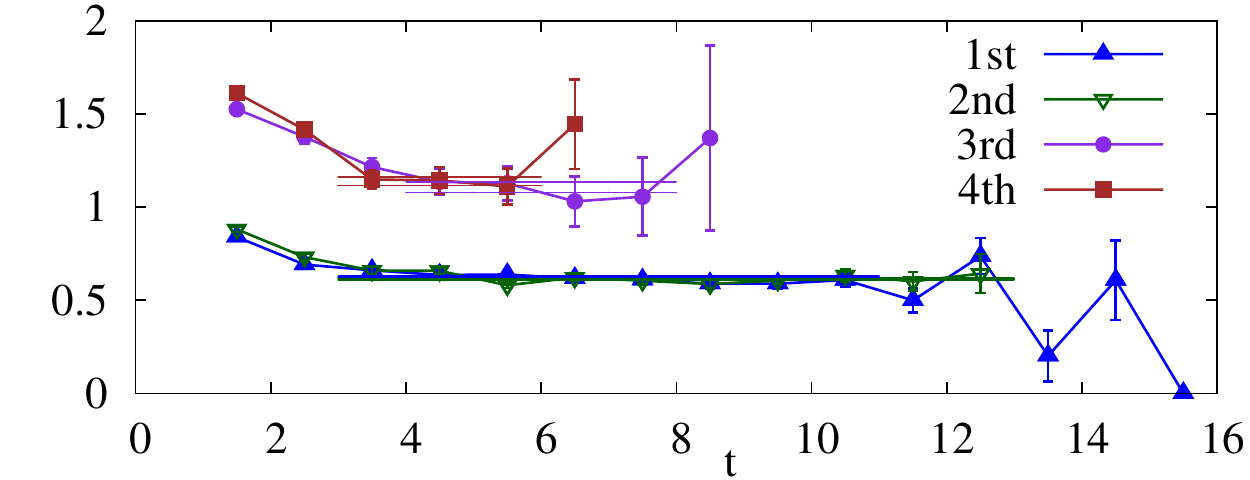}\hspace*{24pt}\hfil\\
  \hspace*{12pt}
  \includegraphics[width=0.45\textwidth]{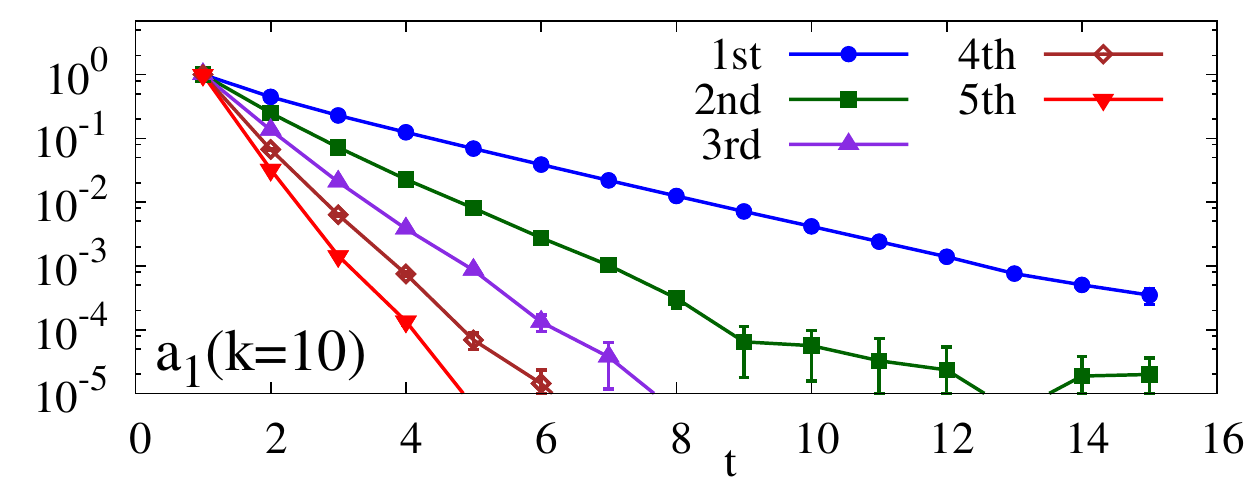}\hfill
  \includegraphics[width=0.45\textwidth]{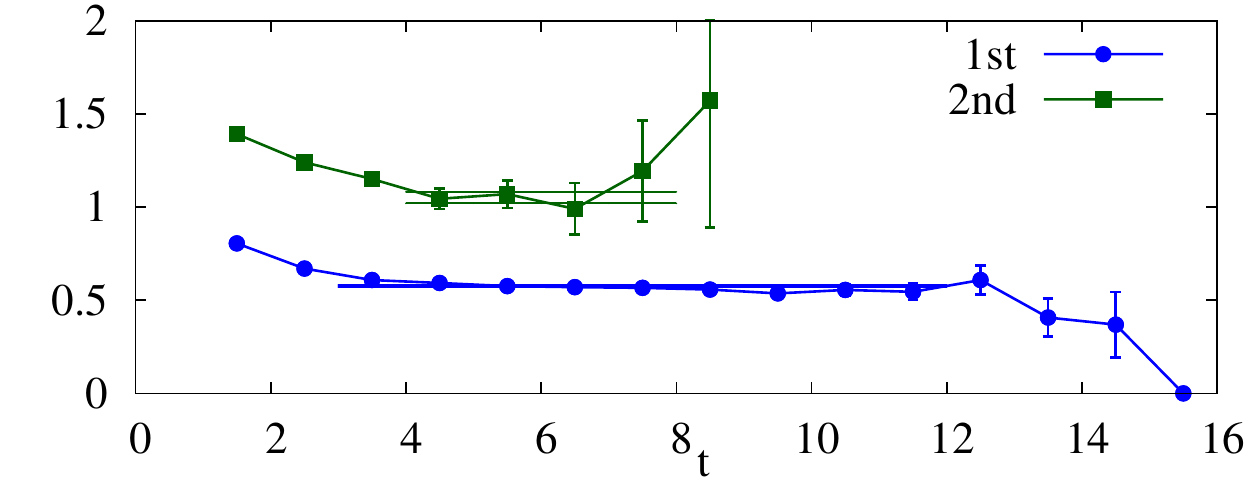}\hspace*{24pt}\hfil\\
  \hspace*{12pt}
  \includegraphics[width=0.45\textwidth]{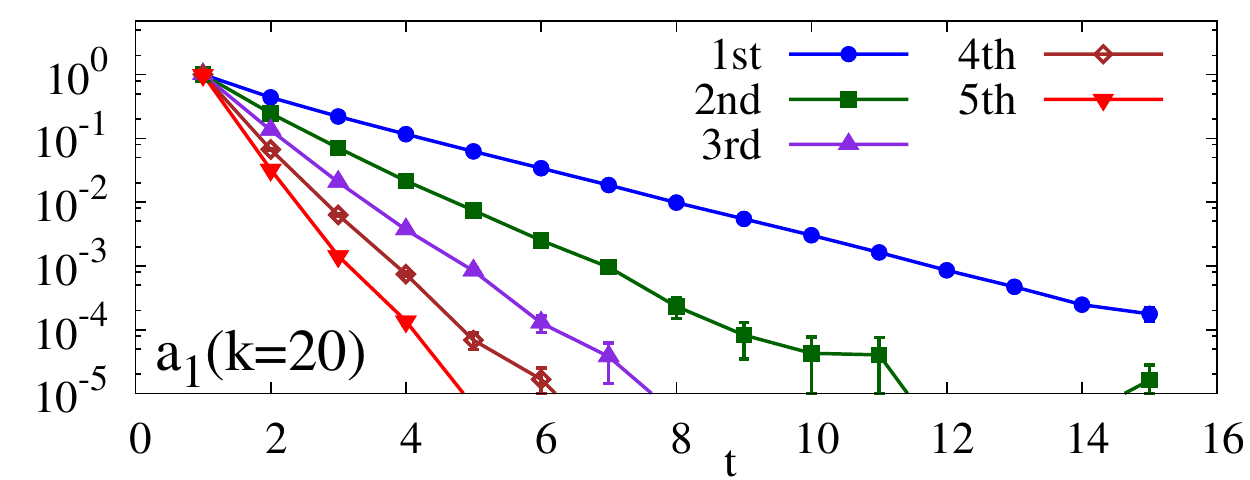}\hfill
  \includegraphics[width=0.45\textwidth]{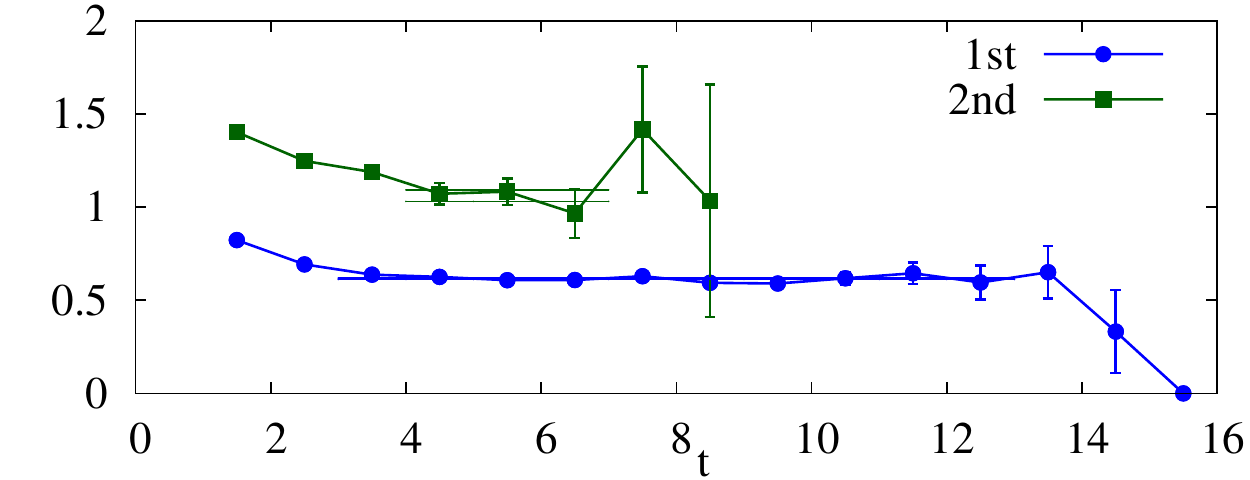}\hspace*{24pt}\hfil\\
  \hspace*{12pt}
  \includegraphics[width=0.45\textwidth]{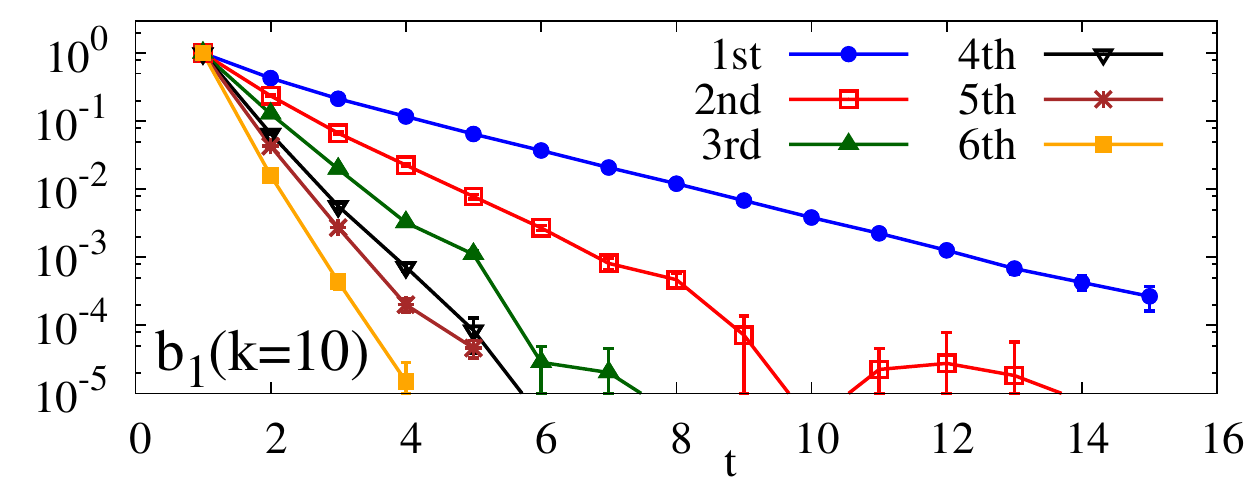}\hfill
  \includegraphics[width=0.45\textwidth]{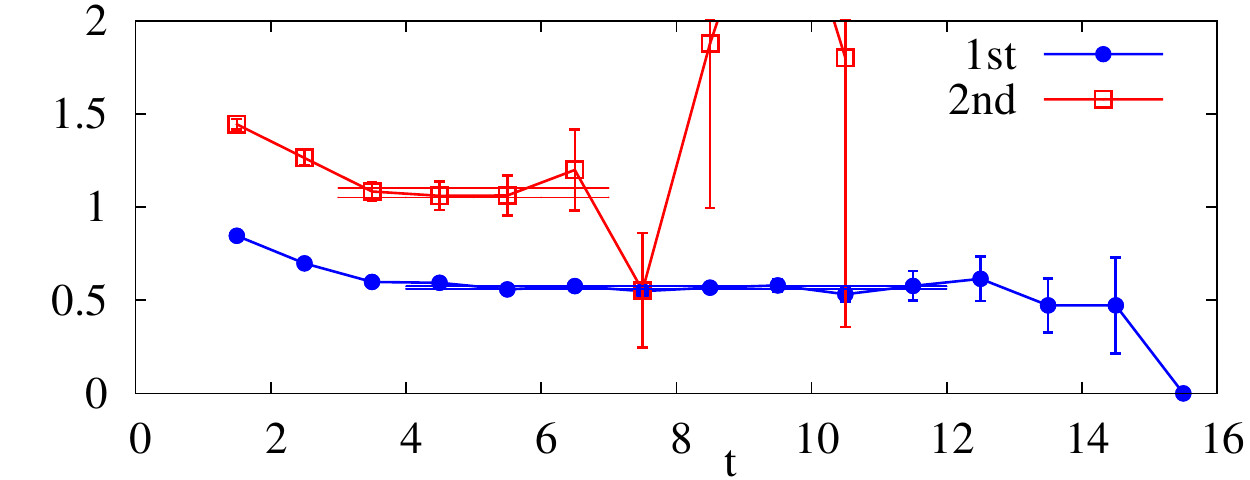}\hspace*{24pt}\hfil\\
  \hspace*{12pt}
  \includegraphics[width=0.45\textwidth]{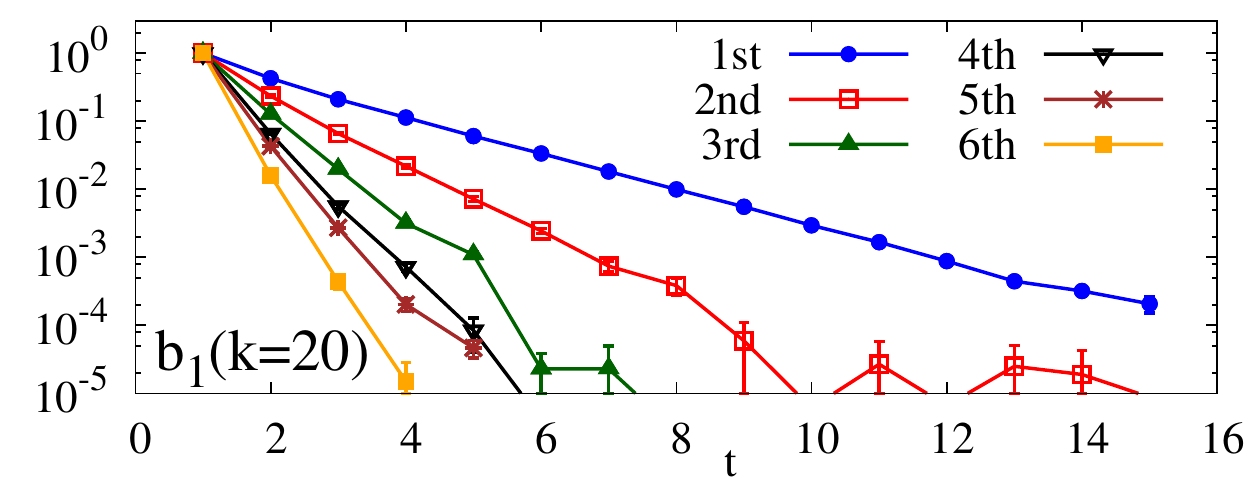}\hfill
  \includegraphics[width=0.45\textwidth]{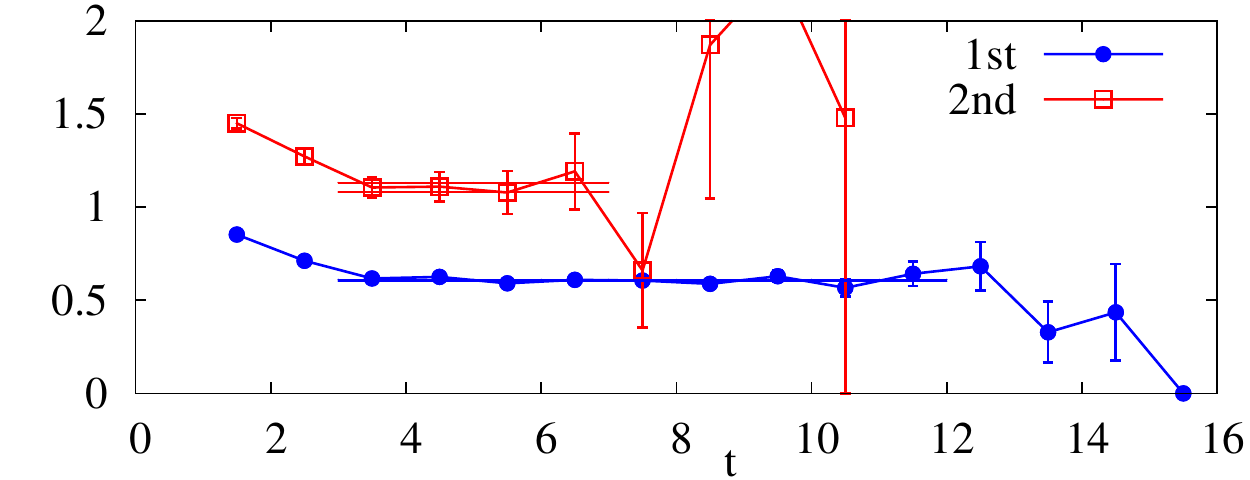}\hspace*{24pt}\hfil\\
   \caption{Upper panel: $\rho$-meson ($J^{PC}=1^{--}$), middle part:
  $a_1$-meson ($J^{PC}=1^{++}$), lower panel:
   $b_1$-meson ($J^{PC}=1^{+-}$).
  We show eigenvalues of the correlation matrix (the left-hand plots)
  and effective masses (the right-hand plots) 
  for $k=10$ and $k=20$  for each meson, respectively.
 Please note the degeneracy of the  levels of the $\rho$ meson, different from the 
  untruncated case.}\label{fig:mesons} 
  \end{figure*} 
   \noindent
  
    \begin{figure}[ht!]
    \includegraphics[width=0.4\textwidth]{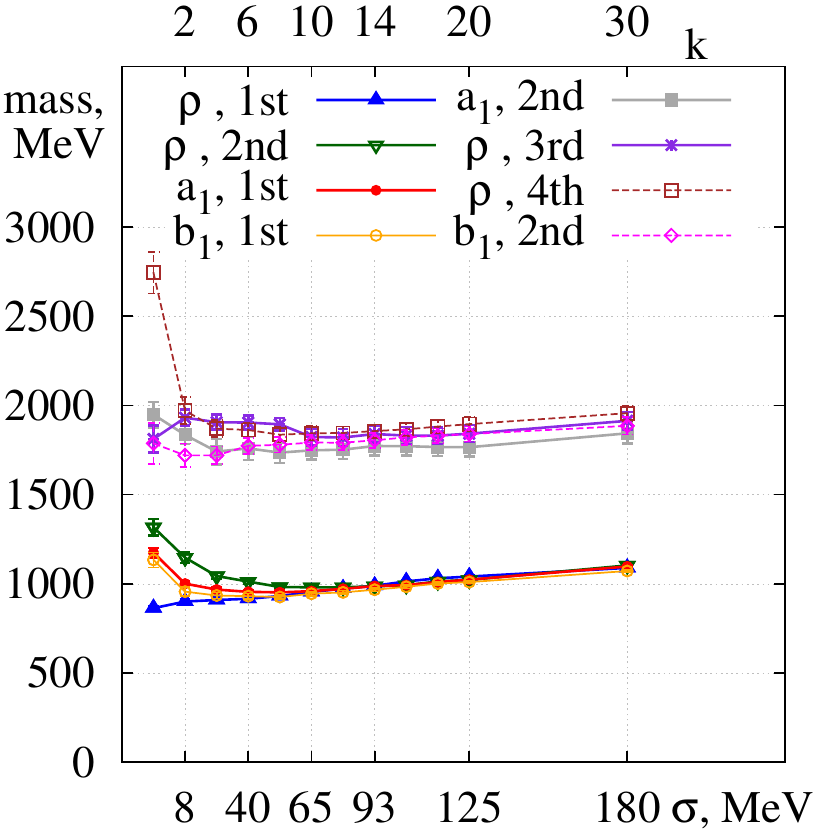}
    \caption{Evolution of hadron masses under the low-mode truncation.
    Both the number $k$ of the removed lowest eigenmodes as well as
    the corresponding energy gap $\sigma$ are given}\label{fig:histo}
  \end{figure} 
  
\begin{table}[t]
\caption{The complete set of $q\bar{q}$ $J=1$ states 
classified according to the chiral basis. 
The symbol $\leftrightarrow$ indicates the states belonging to 
the same representation $R$ of $SU(2)_L \times SU(2)_R$
that must be degenerate in the $SU(2)_L \times SU(2)_R$ symmetric world.}
\begin{ruledtabular}
\begin{tabular}{cc}
$R$  & mesons\\
\hline
$(0,0)$&$\omega(I=0,1^{--}) \leftrightarrow f_1(I=0,1^{++})$\\
$(1/2,1/2)_a$&$\omega(I=0,1^{--}) \leftrightarrow b_1(I=1,1^{+-})$\\
$(1/2,1/2)_b$&$h_1(I=0,1^{+-}) \leftrightarrow \rho(I=1,1^{--}) $\\
$(0,1) \oplus (1,0)$&$a_1(I=1,1^{++} )\leftrightarrow \rho(I=1,1^{--})$\\
\end{tabular}\label{tab:t1}
\end{ruledtabular}
\end{table}

Before discussing  the symmetry issue we want to exclude a possibility that
the observed energy levels are  levels of a free quark-antiquark pair
in a box. For a system of a free quark and antiquark the lowest level
would represent the energy of the quark and antiquark with the $p=0$ 
momentum, which is compatible with the $S-$ and incompatible with the $P$-wave
of relative motion. The next level would correspond to $p=2\pi/L$ ($L$ is the
spatial box size)
 for each quark and is compatible with both $S$- and $P$-waves. Consequently,
a system of a free quark and antiquark in a box cannot produce  degenerate
ground states of opposite parities.
We do observe levels of a  nontrivially bound (confined) quark-antiquark system.

\paragraph{4. Symmetries.}
Which symmetries does  the observed degeneracy represent? All possible
$SU(2)_L \times SU(2)_R$ multiplets of the $J=1$ mesons are listed in
Table \ref{tab:t1} \cite{G3,G2}.

When chiral symmetry is restored but the states still exist we expect
that all  mesons will fall into multiplets of the chiral
group  and within each independent chiral multiplet the mesons will
be degenerate. In the chirally
symmetric world there must be two independent $\rho$-mesons. One of them
is a member of the $(0,1) \oplus (1,0)$ multiplet and can be created
only by an operator with the same chiral structure, e.g., by the vector current,
and that should
be degenerate with its chiral partner $a_1$. Another $\rho$-meson forms together
with its chiral partner $h_1$  the $(1/2,1/2)_b$ representation and is coupled
only to an operator of the type $\bar q \sigma^{0i} \vec \tau q$.

A degeneracy of  the $(0,1) \oplus (1,0)$ $\rho$-meson with the $a_1$ meson
is a clear signal of the chiral $SU(2)_L \times SU(2)_R$ restoration.
We have not yet studied the isoscalar mesons, since
they contain  disconnected graphs and their observation represents a
challenge for the simulation. An observation of the degeneracy within one
of the multiplets is sufficient to confirm chiral restoration, however.
Consequently, a
similar degeneracy should be seen in other chiral pairs.
%Still it would be very interesting to measure in future the isoscalar
%mesons as well.

The $U(1)_A$ symmetry connects the $(1/2,1/2)_b$ $\rho$-meson with the
$b_1$ meson \cite{G3,G2}.  At a  truncation energy around 50  MeV the
onset of this degeneracy 
is also seen. We conclude that
simultaneously both  $SU(2)_L \times SU(2)_R$ and $U(1)_A$
symmetries get restored \footnote{In our previous study with the Wilson-type
Dirac operator  \cite{GLS} we claimed a non-restoration of the $U(1)_A$ 
symmetry. We attributed the reason to the unclear contribution of zero modes in that
not exactly chirally symmetric Dirac operator. Given our present results obtained 
with the overlap Dirac operator, we have revisited our old analysis of the $b_1$
meson and traced an inadequate choice of the fit interval. Upon reanalysis
the result is consistent to the present one.}.
Both chiral and $U(1)_A$ breaking 
are produced by the same low-lying modes of the Dirac operator which
is consistent with the  instanton-induced mechanism
\cite{Hooft,Shuryak,Diakonov}.
 
The restored $ SU(2)_L \times SU(2)_R \times U(1)_A$ symmetry requires a
degeneracy of four mesons that belong to $(1/2,1/2)_a$ 
and $(1/2,1/2)_b$ chiral multiplets \cite{G3,G2}, see Table \ref{tab:t1}.  
This symmetry does not require, however, a degeneracy of these four states with other 
mesons, in particular with $a_1$ and its chiral partner $\rho$. We clearly
see the latter degeneracy. This implies that there is some higher symmetry, 
that includes 
$ U(2)_L \times U(2)_R $ as a subgroup. While a degeneracy
of all isovector mesons is sufficient to claim this, this higher symmetry
requires also a degeneracy of all mesons in Table \ref{tab:t1}, which is a task for
future studies. 

One could ask whether this
situation is similar or not to the high temperature deconfining regime, where 
chiral and $U(1)_A$ symmetries are also restored. In the latter case
correlators of the chiral partners become also identical. This identity does 
not imply, however, that there is a complete spectrum of the  quark-antiquark
bound ground and excited states. The high temperature affects the gluodynamics.
In our case, in contrast, the gluodynamics is kept intact. The valence quarks
have no effect on the Polyakov and 
Wilson loops as well as on the potential between the static
infinitely heavy quarks.

\paragraph{5. Our interpretation. The QCD dynamical string.}
 It was suspected for a long time
and illustrated within the abelian models that the colored quarks are connected
by the color-electric flux tube which, if long enough, can be approximated by a
string. This picture has obtained some support from the lattice simulations
with the static quark sources, where a flux tube as well as the linear
potential have been  observed (for review and references see \cite{Bali}). That
string is nondynamical (in the sense that its ends are fixed), 
however, and can only illustrate to some
extent  physics of hadrons made of heavy quarks. In the light quark
sector the fast motion of quarks at the ends of a possible string, the
respective chiral symmetry as well as its dynamical breaking should be of
great importance. In this case it is even apriori  unclear whether the stringy
picture has something to do with reality. There is no consistent theory
of the QCD string with quarks at the ends. 

The energy of the string with unbroken chiral symmetry
is stored in the gluonic
color-electric flux tube, that is created by the color charges of  quarks at
the ends of the string. This energy should not depend on the orientation
of the quark spins, because the quark spins can interact only with the
color-magnetic fields \cite{G1}. Consequently, one expects that the energy
of the system at a given $J$  
should be the same for all possible orientations of the quark spins and their
parities.

All eight mesons in Table \ref{tab:t1} can be combined into a reducible chiral
representation \cite{CJ}:
\begin{eqnarray}
[(0,1/2)+(1/2,0)] \times [(0,1/2)+(1/2,0)] =\hspace{15mm} \nonumber\\
(0,0) + (1/2,1/2)_a +(1/2,1/2)_b + [(0,1)+ (1,0)]\;.\hspace{4mm}
\end{eqnarray}
They exhaust all possible chiralities of quarks and antiquarks, i.e.,
their spin orientations, 
as well as possible spatial and charge parities for non-exotic mesons.
The observed degeneracy of all these mesons suggests that we see the
energy levels of the dynamical QCD string that connects the ultra-relativistic quark
and antiquark  with the total spin $J=1$.

In Fig. \ref{fig:histo} we show the two lowest energy levels. 
We actually get also a signal of the third level, as can be seen
from the eigenvalues of the correlation matrices on Fig.  \ref{fig:mesons}.
The quality the third level is not sufficient
to show this level in Fig. \ref{fig:histo}.

The observed radial levels at the truncation energy, at which we see the onset of the symmetry,
are approximately  equidistant  
and are compatible with the simple relation
\begin{equation}
E_{n_r} = (n_r +1)\hbar \omega,~~~ n_r=0,1,...
\end{equation}
The extracted value of the fundamental
string excitation quantum at the truncation energy 65 MeV ($k=10$) 
amounts to $ \hbar \omega = (900\pm70)$ MeV \footnote{Approximating the observed levels
with the quadratic relation $E_{n_r}^2 \sim  (n_r+1)$ leads to significantly larger deviations. 
The excited levels might be strongly affected by the finite lattice volume. Consequently,
simulations with larger volumes are required in order to exclude or establish a linear or quadratic dependence.}.
In order to include into this quantization law the rotational levels
one should  study  mesons with $J\neq 1$, which is planned.

\paragraph{6. Is this string of the Nambu-Goto type?}
A principal result
for the classical Nambu-Goto open string  
is that the energy of the string is described in terms of its orbital angular
momentum $L$ (which is a conserved quantum number): $M^2  \sim L$.
For the string with chiral quarks at the ends  
the orbital angular momentum $L$ of the relative motion of the
quark and antiquark is not a conserved quantum number, though \cite{GN}.
For instance, two independent $\rho$-mesons at the same energy
level are represented by the mutually orthogonal fixed superpositions of the
$S$- and $D$-waves.
\noindent
\begin{eqnarray}
\displaystyle |(0,1)+(1,0);1 ~ 1^{--}\rangle&=&\sqrt{\tfrac23}\,|1;{}^3S_1\rangle+\sqrt{\tfrac13}\,|1;{}^3D_1\rangle,\nonumber\\
\displaystyle |(1/2,1/2)_b;1 ~ 1^{--}\rangle&=&\sqrt{\tfrac13}\,|1;{}^3S_1\rangle-\sqrt{\tfrac23}\,|1;{}^3D_1\rangle.\nonumber
\end{eqnarray}
Hence a description of a dynamical string with chiral quarks at the
ends is impossible in terms of a fixed $L$. The total angular momentum $J$ is a
conserved quantum number, of course, as required by Poincare-invariance.

\acknowledgments
We are deeply grateful to S. Aoki, S. Hashimoto and T. Kaneko for their
suggestion to use the JLQCD overlap gauge configurations and quark propagators,
for their help and hospitality during our visit to KEK. Support from the
Austrian Science Fund (FWF) through the grants DK W1203-N16 and 
P26627-N16 is acknowledged.

\end{document}